\documentclass[nofootinbib,aps,prd,groupedaddress,preprintnumbers,%
  showpacs,showkeys,floatfix,amssymb,amsfonts]{revtex4-1}  
\usepackage{hyperref}
\usepackage[usenames]{color}
\usepackage{todonotes}
\usepackage{bbold}
\usepackage{amsmath}
\usepackage{multirow}
\usepackage{graphicx}
\usepackage{xfrac}
\usepackage[utf8]{inputenc}
 
\usepackage{amsfonts}
\usepackage{amssymb}
\usepackage{extarrows}
\usepackage[utf8]{inputenc}
\usepackage[normalem]{ulem}

\usepackage[sort&compress]{natbib}

\newcommand{\be}{\begin{equation}}
\newcommand{\ee}{\end{equation}}
\newcommand{\bea}{\begin{eqnarray}}
\newcommand{\eea}{\end{eqnarray}}

\let\oldbibliography\thebibliography
\renewcommand{\thebibliography}[1]{\oldbibliography{#1}
\setlength{\baselineskip}{9.5pt}
\setlength{\itemsep}{5pt}} %Reducing spacing in the bibliography.
\begin{document}
\title{Mellin Moments of the Unpolarized Gluon PDF in the Proton \\[1ex] from Nonlocal Operators in Lattice QCD}
\author{
%\vspace*{0.35cm}
  Joseph Delmar$^{1,2}$,
 Krzysztof Cichy$^3$,
  Martha Constantinou$^{1}$,
 Yong Zhao$^{2}$ %\\[2ex]
}

\affiliation{
  \vskip 1 cm
    $^1$Department of Physics, Temple University, 1925 N. 12th Street, Philadelphia, PA 19122-1801, USA\\ 
  \vskip 0.05cm
  $^2$ Physics Division, Argonne National Laboratory, Lemont, IL 60439, USA\\
  \vskip 0.05cm
  $^3$ Faculty of Physics and Astronomy, Adam Mickiewicz University, Uniwersytetu Pozna\'{n}skiego 2, 61-614 Pozna\'{n}, Poland \\
  \vskip 0.05cm
  }

\begin{abstract}
We present a lattice QCD determination of the Mellin moments of the unpolarized gluon parton distribution function in the proton. The analysis is based on matrix elements of nonlocal gluon operators coupled to momentum-boosted proton states. The calculation relies on an $N_f=2+1+1$ ensemble of maximally twisted mass fermions with clover improvement and the Iwasaki-improved gauge action, at a pion mass of approximately 260~MeV. Working within the short-distance operator product expansion (OPE) of the reduced gluon Ioffe-time distribution, we extract ratios of higher-order gluon moments, $\langle x^n\rangle$ with $n>1$, to the gluon momentum fraction, $\langle x\rangle$.
We investigate systematic effects associated with the truncation of the order of moment in the OPE, the choice of minimum and maximum Wilson-line separations entering the analysis, and the treatment of mixing with the quark-singlet under perturbative matching. The stability of the extracted moments is further studied under scale evolution using DGLAP equations, allowing us to assess uncertainties related to perturbative truncation by varying the scale. Our work provides a determination of the ratio $\langle x^3\rangle_g/\langle x\rangle_g$ at a scale of 2~GeV, with uncertainties that account for both statistical and the dominant theoretical systematic uncertainties.

\end{abstract}

\maketitle

\section{Introduction}
\label{sec:intro}

Understanding the gluonic structure of hadrons is a central goal of modern hadronic physics. As carriers of the strong force, gluons play a dominant role in determining the mass, momentum, and internal dynamics of hadrons. Despite their importance, quantitative knowledge of gluon parton distribution functions (PDFs) remains less precise than for their quark counterparts, with some recent phenomenological progress reported in Refs.~\cite{Hou:2019efy,Moffat:2021dji,Ball_2022}. Reliable first-principles calculations of gluon PDFs are therefore essential to support current and future experimental programs in major facilities obtained at high-energy colliders such as the Large Hadron Collider (LHC) and the planned Electron-Ion Collider (EIC)~\cite{accardi2014electronioncolliderqcd,Abdul_Khalek_2022}. 

Lattice QCD provides a nonperturbative framework for investigating hadron structure from first principles. However, the light-cone nature of PDFs prevents their direct computation in Euclidean spacetime. Over the past decade, several approaches have been developed to overcome this limitation by relating Euclidean matrix elements of nonlocal operators to light-cone PDFs through effective theory expansion or factorization. Among these, the frameworks of Large Momentum Effective Theory~\cite{Ji:2013dva,Ji:2014gla,Ji:2020ect} (quasi-PDFs), short-distance factorization (pseudo-PDFs~\cite{Radyushkin:2017cyf,Radyushkin:2018cvn} and current-current correlators~\cite{Braun:2007wv,Ma:2014jla,Ma:2017pxb}), hadronic tensor~\cite{Liu:1993cv}, and Compton amplitudes with operator product expansion (OPE)~\cite{Detmold:2005gg,Chambers:2017dov,Detmold:2021qln} have enabled access to the $x$-dependence of parton distributions and their higher Mellin moments using matrix elements of boosted hadrons. Additionally, smearing~\cite{Davoudi:2012ya} and gradient-flow~\cite{Shindler:2023xpd} methods have been introduced to directly calculate Mellin moments beyond the orders which are traditionally limited by power-divergent operator mixings. These methods have been successfully applied to a variety of quark distributions (see, e.g., reviews in Refs.~\cite{Cichy:2018mum,Ji:2020ect,Constantinou:2020hdm,Lin:2025hka,Proceedings:2026xrb}) and, to a lesser degree, to gluonic contributions~\cite{Fan:2018dxu,Zhang:2018diq,Fan:2020cpa,Fan:2021bcr,Salas-Chavira:2021wui,HadStruc:2021wmh,HadStruc:2022yaw,Khan:2022vot,Fan:2022kcb,Chowdhury:2024ymm,GOOD2026140067, chen2025unpolarizedgluonpdfnucleon}. 

In a recent work, we presented a lattice QCD determination of the $x$-dependent unpolarized gluon PDF of the proton using the pseudo-distribution approach, including, for the first time, the elimination of mixing with the quark-singlet contribution using lattice QCD-computed quark matrix elements~\cite{Delmar:2023agv}. That study demonstrated that high-statistics lattice QCD data for nonlocal gluon operators can be combined with perturbative matching to obtain meaningful constraints on the gluon PDF, with results consistent with phenomenological global analyses within uncertainties. The analysis relied on a reconstruction of the $x$-dependence, which necessarily introduces assumptions about the functional form and is sensitive to the limited range of accessible Ioffe time.

An alternative way to characterize partonic structure is through Mellin moments of distribution functions. Mellin moments encode integral properties of the distribution and are directly related to matrix elements of local operators in the operator product expansion (OPE). In particular, higher Mellin moments provide sensitivity to the large-$x$ behavior of PDFs, which is difficult to constrain experimentally and remains one of the major sources of uncertainty in global analyses. From a phenomenological perspective, higher Mellin moments of the gluon PDF play a critical role in constraining global analyses, where they enter as integral constraints that help stabilize fits and reduce uncertainties. Such moments are difficult to access directly from experimental data and remain among the least well-determined quantities in gluon phenomenology. 
From the lattice QCD perspective, moments offer a more direct connection between nonlocal matrix elements at short distances and light-cone physics, reducing reliance on full functional reconstruction. While Mellin moments can, in theory, be obtained by integrating an $x$-dependent PDF, such integration is affected by several systematic uncertainties, for instance, the inability to reliably access the small- and large-$x$ regions. However, their extraction from short-distance lattice QCD matrix elements offers important advantages. In particular, Mellin moments can be related to specific regions of the reduced Ioffe-time distribution and are therefore less sensitive to assumptions in reconstructing the full $x$-dependence.

In this work, we determine the Mellin moment $\langle x^3 \rangle_g$ of the unpolarized gluon PDF in the proton using lattice QCD. Focusing on the short-distance behavior of the reduced gluon Ioffe-time distribution and its OPE, we extract ratios of higher-order gluon moments to the gluon momentum fraction. We perform a detailed study of systematic effects relevant to the extraction, including the truncation  of the order of moment in the OPE, the choice of Wilson-line separations entering the analysis, perturbative matching and scale evolution, and the treatment of quark-gluon mixing. The stability of the extracted moments under variations of these inputs is examined to assess the reliability of the results.

This paper is organized as follows. In Sec.~\ref{sec:methods}, we outline the theoretical framework underlying the extraction of gluon Mellin moments from matrix elements of nonlocal operators through the OPE. Sec.~\ref{sec:setup} describes the lattice QCD setup and analysis strategy. The numerical results and a discussion of systematic uncertainties are presented in Sec.~\ref{sec:results}, along with a comparison to results from global fits. We conclude in Sec.~\ref{sec:conclusions} with a summary and an outlook.

\section{Methodology}
\label{sec:methods}
\subsection{OPE of the nonlocal gluon operator}
\label{sec:methods_A}

The extraction of Mellin moments from lattice QCD relies on the short-distance behavior of matrix elements containing nonlocal operators. 
In the present work, we consider gauge-invariant nonlocal gluon operators separated by a spatial distance $z$, evaluated between boosted proton states. 
At sufficiently small separations, these operators allow a systematic expansion in terms of local twist-two operators through the operator product expansion (OPE), providing a direct connection between lattice QCD-accessible quantities and moments of parton distribution functions.
Specifically, we consider the Euclidean nonlocal matrix element
\begin{equation}
    \label{eq:gluon_oper}
    M_{\mu i; \nu j}(P,z)
    =
    \langle N(P)|
    \mathrm{Tr}\!\left[
    F_{\mu i}(z) W(z, 0)
    F_{\nu j}(0)  W(0,z)
    \right]
    |N(P) \rangle \,,
\end{equation}
where $F_{\mu\nu}$ is the gluon field-strength tensor, $W$ denotes a straight Wilson line of length $z$ in the fundamental representation, $P$ is the momentum boost of the proton state chosen to be along a single spatial direction, and the trace is taken over color indices. 
More details for the extraction of the matrix elements can be found in Sec.~\ref{sec:setup}.
At small spatial separations, the nonlocal operator product can be expanded through the OPE~\cite{Wilson_OPE_paper} as shown in Ref.~\cite{Izubuchi:2018srq},
\begin{equation}
    \label{eq:gluon_OPE}
    F_{\mu i}(z) W(z, 0) F_{\nu j}(0)  W(0,z) = \sum_{n=0}^\infty \frac{z^{\rho_1}\dots z^{\rho_n}}{n!}C_n(z^2,\mu^2)\mathcal{O}_{\mu i;\nu j,\{\rho \}}^{g,(n)}(\mu)\,,
\end{equation}
where $C_n$ are perturbatively calculable Wilson coefficients and
\begin{equation}
    \label{eq:local_gluon_def}
    \mathcal{O}_{\mu i;\nu j,\{\rho \}}^{g,(n)}(\mu)=\mathrm{Sym}\big\{\mathrm{Tr}[F_{\mu i}(iD_{\rho_1})\dots(iD_{\rho_n})F_{\nu j}]\big\}-\mathrm{traces}\,,
\end{equation}
are local, symmetric, traceless twist-two gluon operators. In Eq.\eqref{eq:local_gluon_def}, the symmetrization is over Lorentz indices and the trace is over color indices. The mixings with local quark operators are suppressed for now. In this formalism, the long-distance behavior of the PDF described by these local operators is separated from the short-distance behavior as shown in Eq.~\eqref{eq:gluon_OPE}.

Taking matrix elements of these operators with boosted proton states, $N(P)$, yields
\begin{equation}
    \langle N(P) | \mathcal{O}_{\mu i;\nu j,\{\rho \}}^{g,(n)}(\mu) | N(P) \rangle = 2 P^{\rho_1}\dots P^{\rho_n} a_{n+2}^g(\mu) \,,
\end{equation}
where $a_n(\mu)=\int_0^1 dx x^{n-1} g(x,\mu)$ are the Mellin moments of the unpolarized gluon PDF. 
Combining Eq.~\eqref{eq:gluon_oper} with Eq.~\eqref{eq:gluon_OPE}
and enforcing the kinematics of the collinear frame results in
\begin{equation}
    M(P,z) = \sum_{n=0}^\infty \frac{(i \nu)^n}{n!}C_n(z^2,\mu^2)\,a_{n+2}^g(\mu)\,,
\end{equation}
where $\nu= - z\cdot P$ is the Ioffe time (here, $P\equiv P^z$). This expression shows explicitly that the short-distance behavior of the nonlocal matrix element generates Mellin moments through a power expansion in $\nu$.

An equivalent perspective is obtained by considering the light-cone gluon Ioffe-time distribution (ITD), which is related to the PDF by a cosine Fourier transform,
\begin{equation}
    \label{eq:ITD_FT}
    \mathcal{I}_{g/S}(\nu,\mu^2) = \int_{0}^1 dx \, \cos(x\nu) xf_{g/S}(x, \mu^2)=\int^1_{0}dx\,\sum_{n=0}^\infty (-1)^n\frac{(x\nu)^{2n}}{(2n)!} x f_{g/S}(x, \mu^2)\,,
\end{equation}
whose Taylor expansion at small $\nu$ directly produces even Mellin moments. 

In the short-distance factorization (SDF) formalism, the light-cone ITD can be extracted from lattice QCD matrix elements by a perturbatively calculable matching relation, which includes the gluon and quark-singlet ($S$) contributions~\cite{Balitsky:2019krf,Balitsky:2021bds}
\begin{equation}
    \label{eq:sdf_matching}
    \begin{split}
   \mathfrak{M}(\nu,z^2)\mathcal{I}_g(0,\mu^2)&= \mathcal{I}_g(\nu, \mu^2)-\frac{\alpha_s N_c}{2\pi}\int_0^1 du \, \mathcal{I}(u\nu,\mu^2) \bigg\{\ln\bigg(\frac{z^2\mu^2 e^{2\gamma_E}}{4}\bigg)\mathfrak{B}_{gg}(u)+4\bigg[ \frac{u+\ln(\bar{u})}{\bar{u}} \bigg]_+ + \frac{2}{3}\bigg[ 1-u^3\bigg]_+ \bigg\} \\[1em]
   & \hspace{1.82cm} -\frac{\alpha_s C_F}{2\pi} \ln\bigg(\frac{z^2\mu^2 e^{2\gamma_E}}{4}\bigg) \int_0^1 dw \, \bigg[ \mathcal{I}_S(w \nu, \mu^2)-\mathcal{I}_S(0,\mu^2) \bigg] \mathfrak{B}_{gq}(w)\,,
   \end{split}
\end{equation}
where $N_c=3$ is the color number. $\mathcal{I}_{g/S}(\nu,\mu^2)$ indicates the ITD of the gluon or quark singlet, $\bar{u}\equiv 1-u$, and $\mathfrak{M}(\nu,z^2)$ is the reduced ratio of matrix elements,
\begin{equation}
    \label{eq:double_ratio}
        \mathfrak{M}_g(\nu,z^2) \equiv \bigg( \frac{M_g(\nu,z^2)}{M_g(\nu,0)|_{z=0}} \bigg) \bigg/ \bigg( \frac{M_g(0,z^2)|_{p=0}}{M_g(0,0)|_{p=0,z=0}} \bigg)\,,
\end{equation}
which depends on the Lorentz-invariant quantities $\nu$ and $z^2$. The matching kernels are
\begin{equation}    \label{eq:B_Lkernel}
    \mathfrak{B}_{gg}(u) = 2\bigg[\frac{(1-u\bar{u})^2}{\bar{u}}\bigg]_+\,,\qquad 
\mathfrak{B}_{gq}(u) =  1 + \bar{u}^2 \,,
\end{equation} 
and the plus prescription is given by
$\int^1_0 [f(u)]_+ \mathfrak{M}_g(u\nu)=\int^1_0 f(u) (\mathfrak{M}_g(u\nu)-\mathfrak{M}_g(\nu))$. 
The reduced ratio in Eq.~\eqref{eq:double_ratio} provides several practical advantages. 
In particular, since the operator is multiplicatively renormalizable, the double ratio cancels the overall renormalization factors, including ultraviolet divergences associated with the Wilson-line self-energy~\cite{Ji:2017oey,Ishikawa:2017faj,Green:2017xeu}. 
It leads to a quantity that depends only on the Lorentz invariants $\nu$ and $z^2$. 
In addition, this normalization can suppress discretization artifacts and higher-twist effects, making the short-distance matching more stable~\cite{Orginos:2017kos}.

Inserting the Taylor expansion of the light-cone ITD of Eq.~\eqref{eq:ITD_FT} into the matching relation of Eq.~\eqref{eq:sdf_matching} relates the Mellin moments directly to the lattice QCD calculable reduced ratio. Evaluating the integrals yields the sum expansion
\begin{equation}
    \label{eq:ME_OPE_matching}
    \begin{split}
        \mathfrak{M}_g(\nu,z^2) \langle x \rangle_g^{\mu^2} = &\sum_{n=0}^\infty (-1)^n \frac{\nu^{2n}}{(2n)!} \langle x^{2n+1} \rangle_g^{\mu^2} \bigg[ 1-\frac{\alpha_s N_c}{2\pi}   \bigg\{2\ln\bigg(\frac{z^2\mu^2 e^{2\gamma_E}}{4}\bigg)\bigg(\frac{11}{12} + \frac{1}{1+n} - \frac{1}{2+n} + \frac{2}{3+n}-H_{4+n} \bigg) \\[0.25em]
        %%%%%%%%%%%%%%%%%%%%%%%%
        &\hspace{2.8cm}+\frac{n}{n+1} - H_n + \frac{H_n^2+H^{(2)}_n}{2}  -\frac{3}{4} + \frac{3}{2+n}-\frac{3}{3+n}+\frac{1}{4+n} \bigg\}\bigg]\\[0.25em]
        %%%%%%%%%%%%%%%%%%%%%%
        &-\frac{\alpha_s C_F}{2\pi}\sum_{n=1}^\infty (-1)^n \frac{\nu^{2n}}{(2n)!}\langle x^{2n+1} \rangle_S^{\mu^2} \, \ln\bigg(\frac{z^2 \mu^2 e^{2\gamma_E}}{4}\bigg) \bigg(\frac{2}{1+2n}-\frac{1}{1+n}+\frac{1}{3+2n} \bigg)\,,
    \end{split}
\end{equation}
where $H_n$ and $H_n^{(2)}$ are the $n^{th}$ harmonic number and generalized harmonic number of order 2, respectively. Eq.~\eqref{eq:ME_OPE_matching} makes explicit how the reduced ratio encodes an infinite tower of odd Mellin moments of the gluon PDF, with perturbative Wilson coefficients that depend on $z^2\mu^2$ and on the moment index $n$.
The second line reflects the quark-singlet contribution entering through quark-gluon mixing in the matching.
The left-hand side of Eq.~\eqref{eq:ME_OPE_matching} is weighted by the gluon momentum fraction. Thus, to extract the gluon PDF, knowledge of the gluon momentum fraction $\langle x \rangle_g^{\mu^2}$ should be available from a separate calculation using local operators (see, e.g. Refs.~\cite{Alexandrou:2017oeh,Alexandrou:2020sml}). Alternatively, one can obtain the gluon PDFs normalized with the gluon momentum fraction. In such a case, the Mellin moments become
\begin{equation}
    \label{eq:ME_OPE_matching_ratio}
        \mathfrak{M}_g(\nu,z^2) = 1+ \sum_{n=1}^\infty \tilde{C}^g_n(z^2,\mu^2) \frac{\langle x^{2n+1} \rangle_g^{\mu^2}}{\langle x \rangle_g^{\mu^2}} -\frac{\alpha_s C_F}{2\pi}\sum_{n=1}^\infty \tilde{C}^S_n(z^2,\mu^2) \frac{\langle x^{2n+1} \rangle_S^{\mu^2}}{\langle x \rangle_g^{\mu^2}}\,,
\end{equation}
where we have shifted the sum for the gluonic contribution by one and simplified the notation by introducing $\tilde{C}^{g/S}_n(z^2,\mu^2)$. In principle, we can extract information on both the gluonic contribution and the quark-singlet term. However, in practice, the quark-singlet component in Eq.~\eqref{eq:ME_OPE_matching_ratio} is very small and difficult to extract from noisy lattice QCD data. Thus, within our uncertainties, we do not expect to reliably constrain both. Instead, we neglect the quark-singlet contribution and extract only the Mellin moment ratio $\langle x^3 \rangle_g/ \langle x \rangle_g$. The possible impact of the omitted quark-singlet term can nevertheless be qualitatively estimated using phenomenological extractions of the moments from global analyses, thereby indicating the associated systematic uncertainty (See Sec.~\ref{sec:mixing}).

\subsection{Renormalization Group Invariance}
\label{sec:RG_theory}

In principle, the reduced-ITD, $\mathfrak{M}_g(\nu,z^2)$ contains no explicit scale dependence and should be RG-invariant, as the renormalization factors are canceled by the double-ratio. This allows us to study uncertainties associated with the truncation order of the perturbative matching~\cite{Gao:2021hxl,Su:2022fiu}. We do so by setting an initial scale $\mu_0$ in the matching and evolving the moments to the final scale $\mu=2$ GeV. By varying $\mu_0$ and comparing the evolved results at the final scale, we gain insight on the size of the uncertainty from the truncated matching, as the moments should agree exactly when the matching is known to all orders. Practically, this is only approximately true as the lattice QCD data have a remnant of scale dependence related to the intrinsic UV scale due to the spatial separation $1/z$. However, we expect these effects to be comparably small and contained within the uncertainties of the data where perturbation theory still applies. 
%A natural choice is $\mu_0 \sim 1/z$ which suppresses the logarithmic contribution in the matching kernel. In this work, we choose a central $\mu_0$ such that $\kappa=\frac{\mu_0^2z^2e^{2\gamma_E}}{4}=1$ and vary $\kappa \in (\frac{1}{\sqrt{2}}, 2)$. 

%Since the lattice QCD data have no intrinsic factorization scale, any scale dependence is introduced in the matching kernel through the renormalization scale $\mu$. Typically, final results are reported at fixed scale by choosing directly $\mu=2$ GeV. One can, however, choose an initial scale $\mu_0$ and evolve the results to the final scale of 2 GeV. This allows for study of perturbative effects due to the truncation of the matching as the fixed-order result should the evolved result at the same final scale. 
The evolution of the Mellin moments is governed by the DGLAP equations,
\begin{equation}
    \label{eq:moment_DGLAP}
    \frac{d}{d\ln(\mu^2)} \begin{pmatrix}
    \langle x^n \rangle^{\mu^2}_S\\
    \langle x^n \rangle^{\mu^2}_g
    \end{pmatrix} = 
    \begin{pmatrix}
        \gamma^{qq} & \gamma^{qg} \\
        \gamma^{gq} & \gamma^{gg}
    \end{pmatrix} \begin{pmatrix}
    \langle x^n \rangle^{\mu^2}_S\\
    \langle x^n \rangle^{\mu^2}_g
    \end{pmatrix}\,,
\end{equation} with
\begin{equation}
\gamma^{ij}_n(\mu^2)
=
\frac{\alpha_s(\mu^2)}{2\pi}\gamma^{ij,(0)}_n
+
\left(\frac{\alpha_s(\mu^2)}{2\pi}\right)^2
\gamma^{ij,(1)}_n,
\qquad i,j \in \{q,g\},
\end{equation}
where we use the two-loop results for the anomalous dimensions from Refs.~\cite{Moch_2004,Vogt_2004} as the matching is available at one-loop order so far. We take $\alpha_s(2\,\mathrm{GeV})=0.293$ and carry out the running at two-loop to align with orders used for the anomalous dimension and matching.
In the absence of the gluon and quark-singlet mixing, Eq.~\eqref{eq:moment_DGLAP} reduces to a first-order differential equation dependent only on the anomalous dimension $\gamma^{gg}$. This is the approximation used to report pure lattice QCD estimates. To estimate the impact of the quark-singlet contribution under mixing, one may instead solve the coupled gluon and quark-singlet evolution equation. In practice, we assess the size of the mixing effect using phenomenological input for the quark-singlet moments from global analysis, as described in Sec.~\ref {sec:mixing}. This provides a qualitative estimate of the shift induced by the quark-gluon mixing.

\subsection{Numerical Approach}
\label{sec:numeric_approach}
To determine the Mellin moments normalized with $\langle x\rangle_g$, we fit the reduced matrix element of Eq.~\eqref{eq:ME_OPE_matching_ratio} to a truncated form of the OPE. In the main analysis, we neglect the quark-singlet contribution and consider only the gluonic part,
\begin{equation}
    \label{eq:gluon_only_system}
    \mathfrak{M}_g(\nu,z^2)-1
    =
    \sum_{n=1}^{\frac{n_{max}}{2}}
    \widetilde{C}^{\,g}_{n}(z^2,\mu^2)
    \frac{\langle x^{2n+1}\rangle_g^{\mu^2}}{\langle x\rangle_g^{\mu^2}}\,.
\end{equation}
Here, we test $n_{max}=2,4,6$, which offers the potential to access up to $\langle x^7\rangle_g$. For a given choice of $(z_{min},z_{max})$, we construct the fit using all combinations of the four nonzero momentum boosts and the separations in the selected window. This yields an over-constrained linear system, since, even for a fixed $z$, there are multiple values of the Ioffe time $\nu=P_z z$ entering the analysis. For each fit window, we arrange the data as a vector of matrix elements,
\begin{equation}
    y_i = \mathfrak{M}_g(\nu_i,z_i^2)-1,
\end{equation}
with $i$ labeling the available $(P_z,z)$ pairs. The corresponding design matrix is built from the perturbative coefficients of the truncated OPE, so that the fit takes the linear form
\begin{equation}
    y_i = \sum_j A_{ij}\, r_j,
    \qquad
    r_j \equiv \frac{\langle x^{2j+1}\rangle_g^{\mu^2}}{\langle x\rangle_g^{\mu^2}}.
\end{equation}
In what is called \textit{fixed-order method}, the entries of \(A_{ij}\) are given directly by the matching coefficients evaluated at \(\mu=2~\mathrm{GeV}\). On the contrary, in the \textit{RG-evolved method}, these entries also include the perturbative evolution from the initial scale $\mu_0$ to the final scale of $2~\mathrm{GeV}$. In both cases, the fit remains linear in the unknown moment ratios.

We also investigated fits using the full covariance matrix, as we expect the data to exhibit significant correlations across both common $P_z$ and common $z$. However, the covariance estimated from the bootstrap samples is too poorly conditioned for a stable generalized least-squares analysis. A Ledoit--Wolf shrinkage test~\cite{LEDOIT2004365} indicates that the optimal estimator is driven to the fully shrunk limit, corresponding effectively to retaining only the diagonal uncertainties. For this reason, we use diagonal weighted least squares throughout.

\section{lattice QCD Setup}
\label{sec:setup} 

The data analyzed in this work are based on the evaluation of matrix elements involving momentum-boosted proton states, $N(P)$, that couple to a nonlocal gluon operator. 
The operator is constructed by two gluon field-strength tensors, $F^{\mu\nu}$, located at two lattice points that are spatially separated, along a single spatial direction, for instance, in the $\hat{z}$ direction by distance $z$. 
In addition, the operator contains two straight Wilson lines, connecting points $0\to z$ and $z\to0$, to ensure gauge invariance. 
The matrix element reads
\begin{equation}
    \label{eq:gluon_oper}
    M_{\mu i; \nu j}(P,z)  = \langle N(P)| F_{\mu i}(z) W(z, 0) F_{\nu j}(0)  W(0,z)|N(P) \rangle \,,
\end{equation}
where $F_{\mu\nu}$ is the gluon field strength tensor defined as
{\small{
\begin{eqnarray}
    F_{\mu\nu} (x) &=& \frac{i}{8 g_0} \bigg[ U_\mu(x) U_\nu(x+a\hat{\mu}) U^\dag_\mu(x+a \hat{\nu}) U^\dag_\nu(x) + U_\nu(x) U^\dag_\mu(x+a\hat{\nu}-a\hat{\mu}) U_\nu^\dag(x-a\hat{\mu}) U_\mu(x-a\hat{\mu}) \nonumber \\
    && \qquad + U^\dag_\mu(x-\hat{\mu}) U^\dag_\nu(x-a\hat{\nu}-a\hat{\mu}) U_\mu(x-a\hat{\nu}-a\hat{\mu}) U_\nu(x-a\hat{\nu}) \nonumber \\
    && \qquad+ U^\dag_\nu(x-a\hat{\nu}) U_\mu(x-a\hat{\nu}) U_\nu(x-a\hat{\nu}+a\hat{\mu}) U^\dag_\mu(x) - h.c \bigg]\,,
    \label{Eq:FST}
\end{eqnarray}
}}
and $g_0$ is the bare coupling constant. 
For the gluon operator, we choose indices combinations leading to matrix elements free of contamination~\cite{Balitsky:2019krf,Balitsky:2021bds} that we denote as as ${\cal O}_4$
\begin{eqnarray}
{\cal O}_4 &\equiv& \frac{1}{2} \sum_i F_{it}(x+z \hat{z}) W(x+z \hat{z},x)  F_{it}(x) W(x,x+z \hat{z}) \nonumber \\ 
&-& \sum_{i<j} F_{ij}(x+n\hat{k})  W(x+z \hat{z},x) F_{ij}(x) W(x,x+z \hat{z})\,, \quad i \ne t \ne z \,. 
%&\equiv& \sum_{i<j} F_{ij}(x+z \hat{z}) W(x+z \hat{z},x) F_{ij}(x) W(x,x+z \hat{z}) \nonumber \\ &-& \sum_i F_{it}(x+z \hat{z}) W(x+z \hat{z},x) F_{it}(x)W(x,x+z \hat{z}) \,.
\end{eqnarray}
This operator has been shown to be free of mixing under renormalization~\cite{Zhang:2018diq,Li:2018tpe}. ${\cal O}_4$ has a non-vanishing vacuum expectation value, which we calculate and subtract. 
As in the continuum, the unpolarized gluon PDF obtained by the lattice operator mixes with the unpolarized quark-singlet PDF. This mixing can be eliminated in the matching formalism, not under renormalization, due to the operator's nonlocal nature.
We take this mixing into account in our analysis using results from the global analysis of Ref.~\cite{Ball_2022}, and we quantify its effects by comparing to results with the mixing neglected. 

The matrix elements of Eq.~\eqref{eq:gluon_oper}, $M_{\mu i; \nu j}$, are extracted from the ratio of the three-point over the two-point functions, that is
\begin{equation}
 R_{\cal O}(t_s,\tau,t_0;P,z) = \frac{C^{\text{3pt}}_{\cal O}(t_s,\tau; P, z)}{C^{\text{2pt}} (t_s; P)}\,\, \stackrel{t_s<\tau<t'} \longrightarrow \,\, \frac{4}{3} \left( \frac{m^2}{4 E} - E \right) M_{\cal O}(t_s; P,z)\,.
  \label{Eq:Ratio}
\end{equation}
The variables $t_s$, $\tau$, and $t_0$ denote the sink, operator insertion, and source time slices, respectively. Without loss of generality, we set $t_0=0$. 
The three-point functions $C^{\text{3pt}}_{\cal O}$ relevant for the gluonic contributions to the proton arise from disconnected diagrams and are constructed from the expectation value of the product of a gluon loop with the proton two-point function. For the unpolarized gluon PDF and its moments, we employ the parity projector $\Gamma_0 \equiv \frac{1}{4}(1+\gamma_0)$ in both the three- and two-point functions.
The ground-state matrix element, $M_{\cal O}\equiv M_{\mu i; \nu j}$, is extracted from the ratio through a plateau fit at sufficiently large source–sink separations $t_s$ and with $\tau$ chosen away from both the source and sink to suppress excited-state contamination. The investigation of excited-state effects is presented in our work on the $x$ dependence of the gluon PDF, using the same raw data~\cite{Delmar:2023agv}~\footnote{Ref.~\cite{Delmar:2023agv} uses 3D smearing in the Wilson line, whereas here we use a 4D one, which yields a small improvement in the signal. However, it does not alter the conclusions on the effect from excited states.}.

\medskip
The calculations are carried out on the $N_f=2+1+1$ twisted-mass ensemble
cA211.30.32, employing clover-improved fermions and the Iwasaki gauge action~\cite{Alexandrou:2018egz,ExtendedTwistedMass:2021gbo}. 
The simulation parameters are tuned such that the pion mass is approximately
$m_\pi = 260$ MeV, the lattice spacing is $a=0.0908$ fm, and the lattice size is $32^3 \times 64$, yielding a spatial extent of $L\simeq 3.0$ fm and $Lm_\pi \approx 4$.  The ensemble parameters are summarized in Table~\ref{tab:params}.

\begin{table}[h!]
\centering
\renewcommand{\arraystretch}{1.2}
\renewcommand{\tabcolsep}{6pt}
\begin{tabular}{| l| c | c | c | c | c  | c | c |}
    \hline
    \multicolumn{8}{|c|}{Parameters} \\
    \hline
    Ensemble   & $\beta$ & $a$ [fm] & volume $L^3\times T$ & $N_f$ & $m_\pi$ [MeV] &
    $L m_\pi$ & $L$ [fm]\\
    \hline
    cA211.30.32 & 1.726 & 0.0908  & $32^3\times 64$  & 2+1+1 & 260
    & 4 & 3.0 \\
    \hline
\end{tabular}
\caption{\small{Parameters of the ensemble used in this work.}}
\label{tab:params}
\end{table}

The matrix elements of gluonic operators are known to suffer from substantial gauge noise, necessitating both high statistics and noise-reduction techniques. To enhance statistics, the correlation functions are computed from multiple source locations per configuration. The ensemble consists of approximately 1,200 thermalized gauge configurations~\cite{Alexandrou:2018egz}, and we analyze 200 source positions on each configuration. This strategy, combined with the significant acceleration provided by the
multi-grid solver~\cite{Clark:2016rdz,Alexandrou:2016izb,Bacchio:2017pcp,Alexandrou:2018wiv},
substantially increases the statistics.

Additional statistical improvement is achieved by exploiting cubic symmetry. We evaluate matrix elements in six equivalent kinematic setups, with both the Wilson line and the momentum boost aligned along the $\pm \hat{x}$, $\pm \hat{y}$, and $\pm \hat{z}$ directions. Averaging over these orientations results in a total number of measurements exceeding one million, as detailed in Table~\ref{tab:stat}.

Since the pseudo-ITD analysis requires several Ioffe time combinations, we consider five values for the momentum boost, that is,
$P = 0,\,0.43,\,0.85,\,1.28,$ and $1.71$ GeV. Each boosted matrix element is normalized by its $P=0$ counterpart. We observe non-negligible correlations between numerator and denominator in the reduced ITD of Eq.~\eqref{eq:double_ratio}. To control these correlations, all matrix elements are computed on identical gauge configurations and source positions.
Excited-state effects are investigated using multiple source-sink separations $t_s$. Importantly, this does not introduce additional computational cost, since disconnected contributions are evaluated at open sink time.

\begin{table}[h!]
\begin{center}
\renewcommand{\arraystretch}{1.8}
\begin{tabular}{c|ccccc}
\hline
$P$ [GeV] & $N_{\rm confs}$ & $N_{\rm src}$ & $N_{\rm dir}$ & $N_{\rm meas}$ \\
\hline
0, 0.43, 0.85, 1.28, 1.71 & 1,134 & 200 & 6 & 1,360,800 \\
\hline
\end{tabular}
\vspace*{0.15cm}
\caption{\small{Statistics for each momentum value.
$N_{\rm confs}$ denotes the number of gauge configurations,
$N_{\rm src}$ the number of source positions per configuration,
$N_{\rm dir}$ the number of equivalent spatial orientations,
and $N_{\rm meas}=N_{\rm confs}\times N_{\rm src}\times N_{\rm dir}$
the total number of measurements.}}
\label{tab:stat}
\end{center}
\end{table}

To mitigate gauge noise further, stout smearing~\cite{Morningstar:2003gk} is applied to the gauge links entering both the gluon field-strength tensor and the Wilson line. We use a smearing parameter $\rho=0.129$~\cite{Alexandrou:2016ekb,Alexandrou:2020sml}, and vary independently the number of smearing iterations applied to the field-strength tensor ($N^{\rm F}_{\rm stout}$) and to the Wilson line ($N^{\rm W}_{\rm stout}$). Specifically, 4D smearing is employed for both the field-strength tensor and the Wilson line, which is an improvement compared to Ref.~\cite{Delmar:2023agv}. In total, 25 combinations of smearing levels are analyzed,
with $N^{\rm F}_{\rm stout},\,N^{\rm W}_{\rm stout}=0,5,10,15,20$.
To improve the overlap with the boosted proton ground state, we additionally implement momentum smearing~\cite{Bali:2016lva} for the three highest boosts, $P=0.85,\,1.28,$ and $1.71$ GeV. This method has proven crucial in reducing noise for boosted hadrons with nonlocal operators~\cite{Alexandrou:2016jqi}. The smearing parameter is optimized at $\xi=0.6$.

\section{Results}
\label{sec:results}

In this section, we present the numerical extraction of Mellin moments of the unpolarized gluon PDF from the reduced gluon Ioffe-time distribution. Our analysis focuses on the ratio $\langle x^3 \rangle_g/\langle x \rangle_g$, with exploratory investigations of higher moments through increasingly truncated forms of the OPE. We first examine the lattice QCD matrix elements and the corresponding reduced ratios that enter the analysis, followed by a detailed study of the stability of the extracted moments under variations in the fit window, truncation order, and perturbative evolution. We also explore alternative ratios entering the fit and conduct a qualitative study of the effect of mixing with the quark-singlet contributions.

\subsection{Matrix Elements} 

\begin{figure}[h!]
    \centering
    \includegraphics[scale=0.425]{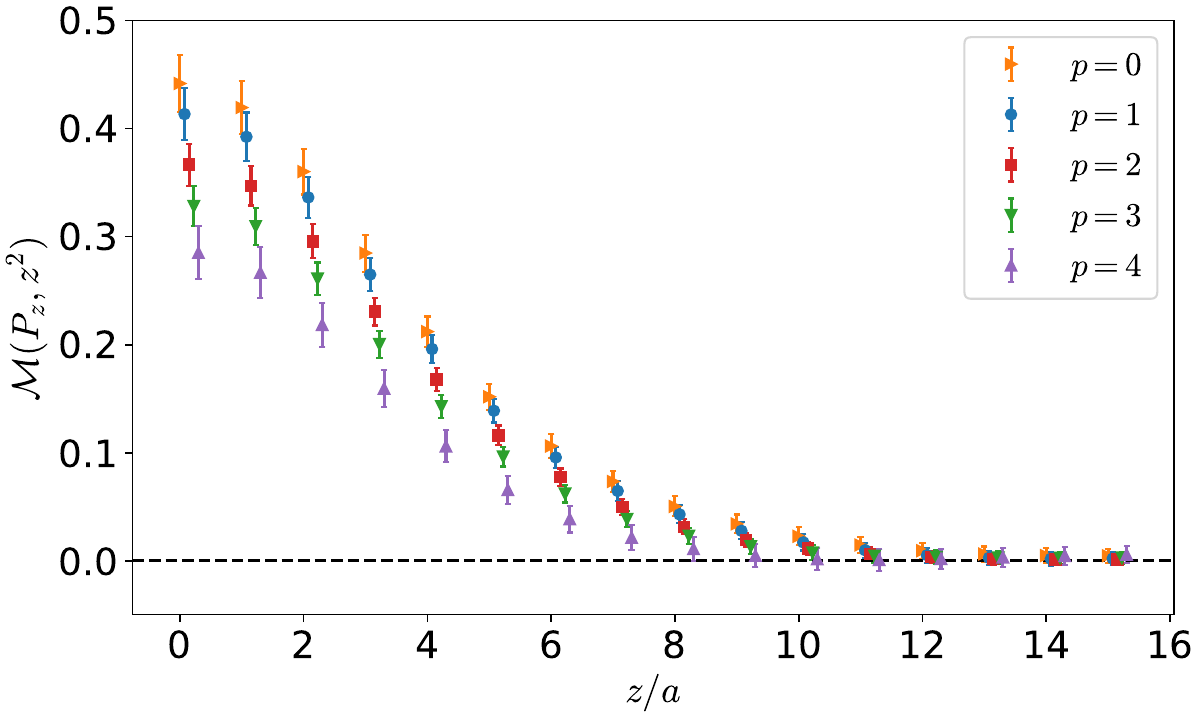} 
        \includegraphics[scale=0.42]{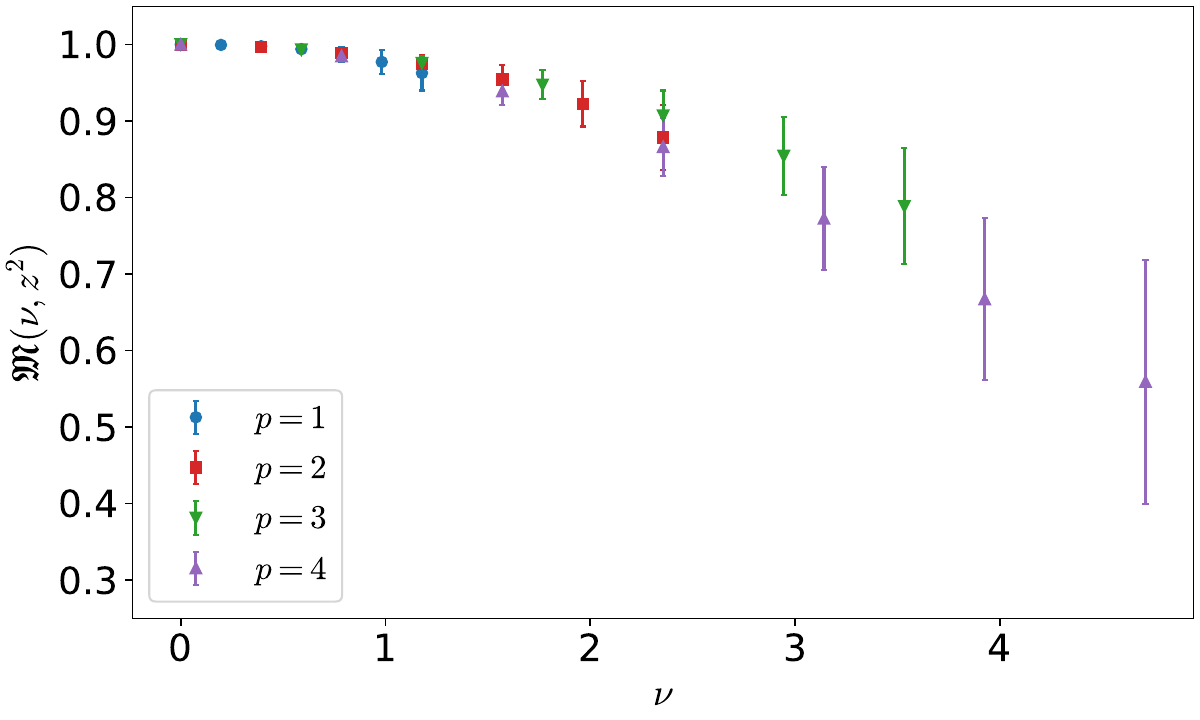}
    \vspace*{-0.5cm}
    \caption{\small{Left: Bare matrix elements as a function of the length of the Wilson line, $z/a$. The data at momentum boost $P=\frac{2\pi}{L} p$ with $p=0,\,1,\,2,3,4$ are shown with orange rightward-pointing triangles, blue circles, red squares, green downward-pointing triangles, and magenta upward-pointing triangles, respectively. The points have been horizontally offset for visibility. Right: Reduced-ITD for all values of $P=\frac{2\pi}{L} p$ and $z$ up to $6a\sim0.54$ fm. The data notation is the same as the left panel.}}    \label{fig:gluon_matrix}
\end{figure}

We begin by presenting the lattice QCD matrix elements entering the analysis, followed by the reduced double ratio constructed from them. These quantities form the basis of the extraction of the gluon Mellin moments through the short-distance expansion discussed in Sec.~\ref{sec:methods_A}. To summarize the bare matrix elements, we compare in the left panel of Fig.~\ref{fig:gluon_matrix} the data for all values of the momentum boost using $t_s=9 a$,  $N^{\rm F}_{\rm stout}=20$, and $N^{\rm W}_{\rm stout}=10$. This combination of parameters is consistent with the optimizations identified in Ref.~\cite{Delmar:2023agv} (see Figs. 2 - 3 therein). In addition, Fig. 6 of Ref.~\cite{Delmar:2023agv} demonstrates stout independence in the double ratio. As anticipated, the signal quality decreases with increasing momentum boost, $P$. For instance, we find that the relative error at $z=0$ for $P=0$ is about 6\%, whereas for $P=1.71$ GeV it is close to 9\% despite the same statistics. 
In all cases, we find that the matrix elements decay to zero at about $z=8a$.
%\KC{We know the decay depends rather on the Ioffe time, not $z$, and indeed only the largest boost becomes consistent with zero at $z=8a$, while the lower ones do so only at much larger $z$. BTW, do we ever make the switch from $z$ as a 4-vector to scalar?}

The double ratio of Eq.~\eqref{eq:double_ratio} is then constructed from the matrix elements, and is shown in the right panel of Fig.~\ref{fig:gluon_matrix}, for the parameters mentioned above. The $\nu$ dependence is reconstructed by all possible combinations of $z$ and $P$, but we constrain $z$ up to $6a\sim0.54$ fm. 
This leads to $\nu_{max}\sim5$. We observe that data points corresponding to different combinations of $(P,z)$ but similar values of $\nu$ largely follow a common trend within statistical uncertainties. At the same time, a residual spread among such points remains visible, particularly at larger values of $\nu$, suggesting the presence of subleading effects beyond the ideal short-distance description. As anticipated, the statistical uncertainties increase toward larger Ioffe times, where both the signal-to-noise ratio deteriorates, and higher-order contributions in the OPE become increasingly difficult to constrain reliably.

\subsection{Fixed-Order Moments}
\label{sec:fixed_order}

We now turn to the results obtained from the matching relation evaluated at a fixed renormalization scale $\mu=2$~GeV, without performing additional renormalization-group evolution. In practice, this means that the perturbative Wilson coefficients entering Eq.~\eqref{eq:ME_OPE_matching_ratio} are computed at $\mu=2$~GeV and inserted directly into the truncated OPE expression, allowing us to extract the moment ratios at that scale in a single step. We refer to this as a \textit{fixed-order} determination, in contrast to the analysis of Sec.~\ref{sec:evolution}, where the moments are first matched at an initial scale $\mu_0$ and subsequently evolved to 2~GeV using DGLAP evolution.
The fixed-order extraction isolates uncertainties arising from choices in the short-distance expansion itself, allowing us to assess the impact of the OPE truncation order and the range of spatial separations included in the fit, independently of other lattice QCD systematics. In this way, we can study the stability of the moment extraction directly at the level of the lattice QCD matrix elements before incorporating perturbative evolution effects. One could also isolate these effects at a scale $\mu_0 \sim 1/z$, but our choice allows for a direct comparison of the extracted moments with PDFs quoted at the conventional scale of 2~GeV.

We emphasize that the OPE is valid at sufficiently small spatial separations, where perturbation theory is applicable. The availability of four non-zero momentum boosts enables us to select a restricted window $(z_{min}, z_{max})$ while maintaining sufficient data to constrain the moments. Even at a fixed value of $z$, using multiple momentum boosts yields an over-constrained system, thereby improving the stability of the extracted moments. A related systematic effect arises from truncating the OPE series in Eq.~\eqref{eq:ME_OPE_matching_ratio}. Although formally an infinite expansion, the higher-order terms decrease rapidly. Therefore, we keep only the leading contributions.  

\begin{figure}[h!]
    \centering
    \includegraphics[scale=0.45]{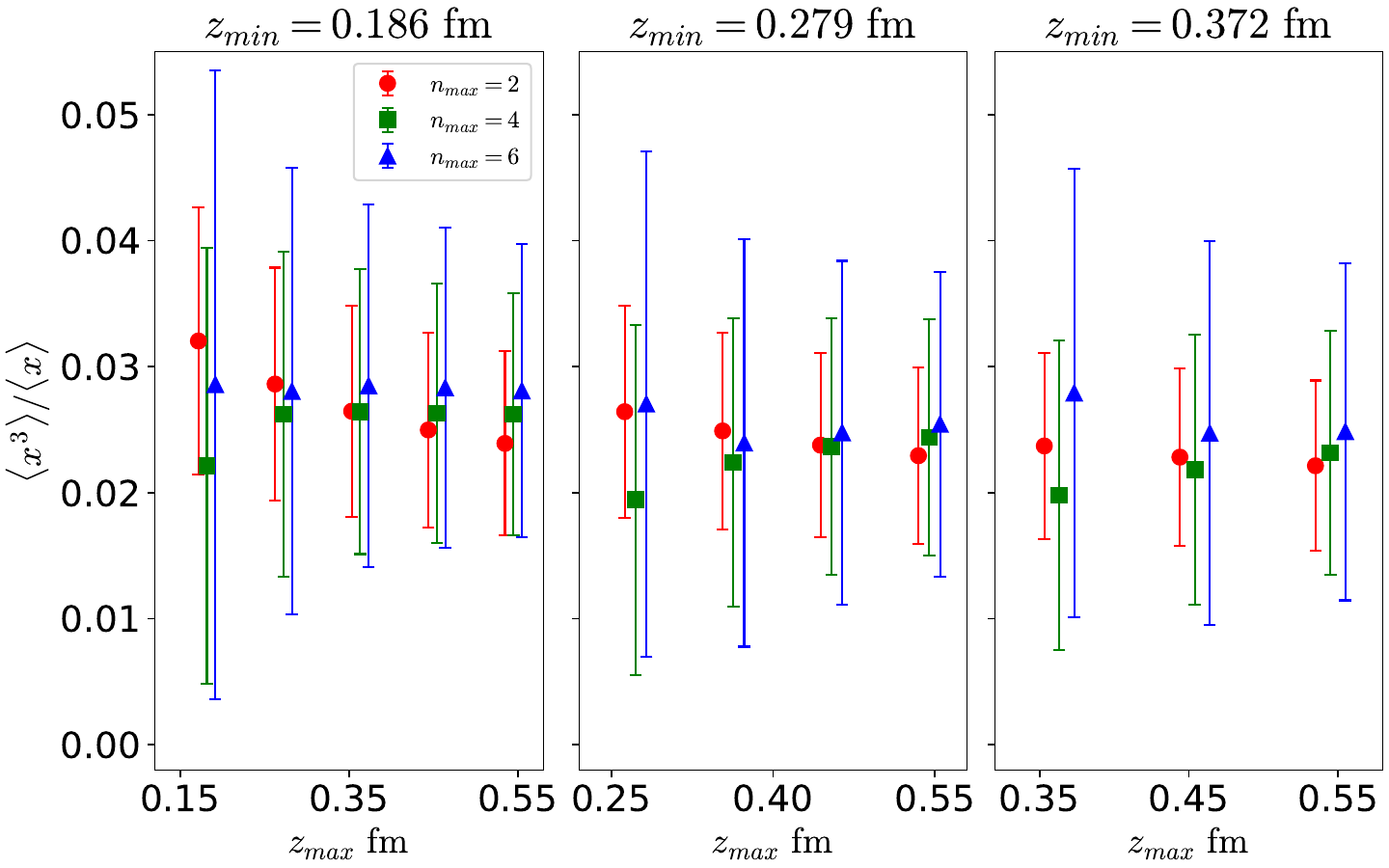}
    %\vspace*{-0.5cm}
    \caption{Fixed-order results for the ratio $\langle x^3 \rangle_g/\langle x \rangle_g$ neglecting the quark-singlet contribution. The panels from left to right are the minimum spatial separation in matrix elements entering the fit ($z_{min}=2a,\,3a,\,4a$ respectively), and the $x$-axis indicates the maximum value of spatial separation entering the fit. The ratios are extracted from the sum in Eq.~\eqref{eq:ME_OPE_matching_ratio} truncated at $n_{max}=2$ (red circles), 4 (green squares), and 6 (blue triangles).}
    \label{fig:no_evo_moments}
\end{figure}

As seen in Eq.~\eqref{eq:ME_OPE_matching_ratio}, the moments are normalized to the leading one, namely, $\langle x \rangle_g$. In Fig.~\ref{fig:no_evo_moments} we present the ratio of $\langle x^3 \rangle_g/\langle x \rangle_g$, demonstrating its dependence on the above mentioned parameters: window of $z$ entering the fit, and truncation of the OPE series, $n_{max}$. In particular, we test all combinations of: \textbf{a)} $z_{min}/a=2,\,3,\,4$, that is $z_{min}=0.182,\,0.272,\,0.363$ fm; \textbf{b)} $z_{max}/a = z_{min},\cdots,6a$; \textbf{c)} $n_{max}=2, 4, 6$, allowing access up to the $\langle x^{(n_{max}+1)} \rangle_g$ moment. Since $z_{min}$ probes the sensitivity to the shortest-distance data points, we exclude $z_{min}=1a$, due to enhanced discretization effects. 
First, as seen across the three panels of Fig.~\ref{fig:no_evo_moments}, increasing $z_{min}$ does not lead to a systematic shift in the central value of the extracted ratio, although the statistical uncertainties increase when fewer points enter the fit. 
Second, varying $z_{max}$ tests higher-twist contamination at larger separations. 
For sufficiently large $z_{max}$, the OPE description may begin to deteriorate due to nonperturbative effects. The data show that the extracted ratio remains stable as $z_{max}$ increases within the considered window, indicating that higher-twist effects are subleading in this region within the reported statistical uncertainties. At the same time, including more $z$ values improves the constraint on the fit when the number of fit parameters ($n$) is kept fixed. As an example, one may compare the decreased statistical errors of $(z_{min},z_{max})=(2a,6a)$ as compared to those from the fit with $(z_{min},z_{max})=(2a,3a)$.
Third, the truncation order of the OPE, controlled by $n_{max}$, tests the convergence of the short-distance expansion. The results obtained with $n_{max}=2,4,6$ are statistically compatible, with no systematic change as higher moments are included. While higher truncation orders allow access to additional moments, the uncertainties grow substantially and these higher moments are not well constrained. This is also related to the fact that $\langle x^5\rangle_g$ and $\langle x^7\rangle_g$ are significantly suppressed compared to the leading term with $\langle x^3\rangle_g$. Thus, we do not show $\langle x^5\rangle_g$ and $\langle x^7\rangle_g$ in this presentation, as they are found to be zero within uncertainties.
Overall, Fig.~\ref{fig:no_evo_moments} demonstrates that, at fixed order, the uncertainties appear to be primarily driven by the number of degrees of freedom in the fit rather than from a particular parameter choice. We conclude that the extracted ratio is stable under reasonable variations of the fit window and truncation order. The central values of the moments agree well across all choices of systematics, and we find that the relative size of $\langle x^3 \rangle$ is between 2-3\% of $\langle x \rangle$. 
We remind the reader that, at this stage, the mixing with the quark-singlet contribution is not considered.

\subsection{Exploration of Mellin Moments from Alternative Ratios}

It was suggested in Refs.~\cite{Gao:2022vyh,Gao:2022uhg} that ratios analogous to Eq.~\eqref{eq:double_ratio} may be constructed using matrix elements at nonzero momentum in the denominator. The main motivation for this choice is a potential reduction in statistical uncertainties by suppressing correlations between boosted matrix elements evaluated at different nonzero values of $P$. In addition, such ratios may partially suppress residual higher-twist effects associated with the leading-twist expansion, although the precise cancellation pattern depends on the choice of selected $P$ values and fit strategy. In the implementations of Refs.~\cite{Gao:2022vyh,Gao:2022uhg}, the denominator momentum was typically chosen such that $P_{\rm denom}\leq P_{\rm numer}$ in order to maintain a controlled hierarchy of scales entering the OPE analysis. 

\begin{figure}[h!]
    \centering
    \includegraphics[scale=0.475]{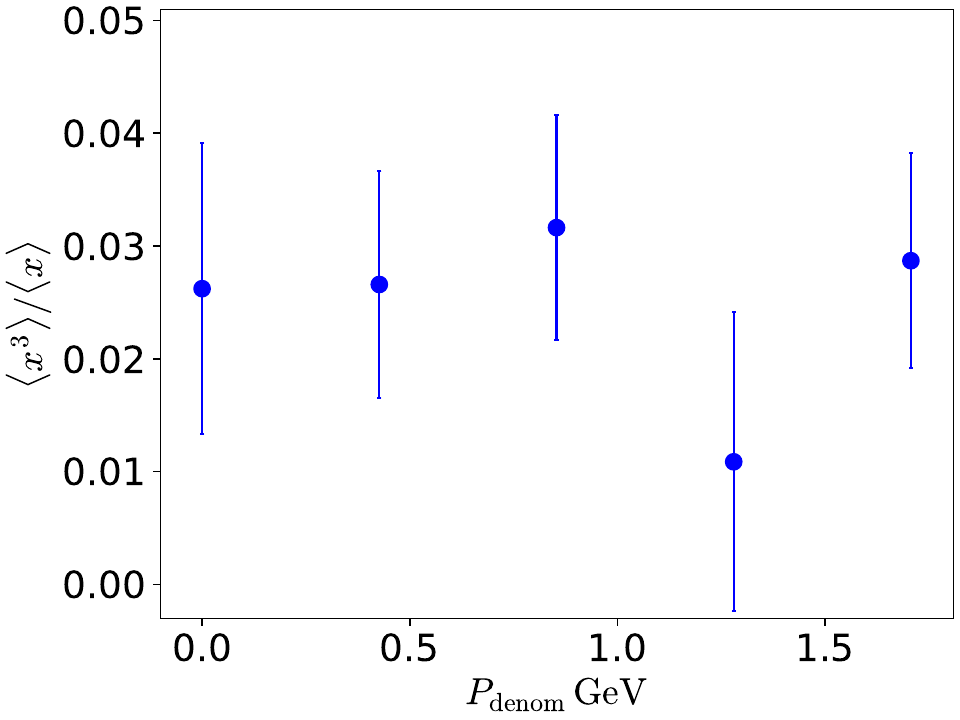}
%    \includegraphics[scale=0.42]{figures/comp_P_denom_rel_err_g_only_no_evo.pdf}
    %\vspace*{-0.5cm}
    \caption{A comparison of ratios $\langle x^3 \rangle_g/\langle x \rangle_g$ for $z\in(2a,3a)$ and $n_{max}{\,=\,}4$ for different values of $P$ in the denominator of the double-ratio. The mixing with the quark-singlet contribution is neglected.}
    \label{fig:ratio_comp_moments}
\end{figure}

Here, we perform an analogous comparison within our fixed-order analysis using the fit range $z\in(2a,3a)$ and truncation order $n_{max}{\,=\,}4$. Unlike the standard reduced ratio of Eq.~\eqref{eq:double_ratio}, the use of a boosted matrix element in the denominator introduces a nonlinear dependence on the fit parameters through the denominator of the OPE expression. Consequently, the extraction no longer yields a linear fit to the Mellin moments. We determine the moments through a direct minimization of a weighted $\chi^2$ function that simultaneously incorporates matrix elements for all $P\neq P_{\rm denom}$ with $P>0$, including all values of $z$ entering the selected fit window.
The resulting moments are shown in Fig.~\ref{fig:ratio_comp_moments}. Overall, we observe consistency within uncertainties for all choices of the momentum in the denominator, except $P_\mathrm{denom} = 1.28\,\mathrm{GeV}$, for which it is difficult to constrain the fit. The value for $\langle x^3 \rangle_g/\langle x \rangle_g$ remains similar using $P_{\rm denom} = 0.43, 0.85, 1.71\,\mathrm{GeV}$.

We emphasize that the analysis with alternative ratios is exploratory rather than a replacement of the standard analysis based on Eq.~\eqref{eq:double_ratio}. Although the nonlinear fit procedure can be implemented consistently using the same underlying data set, it leads to a qualitatively different analysis strategy compared with our linear extraction. Furthermore, while perturbative evolution could also be incorporated in this framework, a systematic study of such implementations lies beyond the scope of the present work. For these reasons, we consider the standard ratio with $P_{\rm denom}=0$, and the results of Sec.~\ref{sec:fixed_order} in what follows.

\subsection{Renormalization-Group Evolution}
\label{sec:evolution}

While the reduced ratio $M_g(\nu,z^2)$ is constructed to be renormalization-group invariant due to the cancellation of ultraviolet factors in the double ratio, the perturbative matching relation of Eq.~\eqref{eq:ME_OPE_matching_ratio} introduces an explicit dependence on the factorization scale $\mu$ through the Wilson coefficients. As outlined in Sec.~\ref{sec:RG_theory}, the moments evolve according to DGLAP, described by Eq.~\eqref{eq:moment_DGLAP}. At finite order in $\alpha_s$, residual scale dependence remains and can be used to estimate uncertainties associated with truncating the perturbative expansion. In particular, we study the behavior of the moments as we gradually evolve from an initial scale $\mu_0$ to the final scale $\mu=2$ GeV. For the running of $\alpha_s$ we use an initial value of $\alpha_s(2\,\mathrm{GeV})=0.293$ and carry out the evolution at two loops according to Ref.~\cite{Stewart_2010}. 
A natural choice for the initial scale $\mu_0$ is suggested by the logarithmic structure of the matching kernel in Eq.~\eqref{eq:ME_OPE_matching}, which contains the term
\begin{equation}
    \ln\!\left(\frac{z^2 \mu_0^2 e^{2\gamma_E}}{4}\right)\,.
\end{equation}
We therefore define the dimensionless parameter
\begin{equation}
    \kappa = \frac{\mu_0^2 z^2 e^{2\gamma_E}}{4}\,,
\end{equation}
such that $\kappa = 1$ corresponds to the scale that eliminates the large logarithm in the matching kernel. This choice corresponds to $\mu_0 \sim 1/z$, which is the expected hard scale governing the short-distance expansion. To assess the sensitivity to perturbative truncation, we vary $\kappa$ in the range $(\frac{1}{\sqrt{2}},\, \sqrt{2})$.
%\KC{Is there a purpose of making $\mu_0\propto\sqrt{\kappa}?$. Yong's papers usually have a linear relation between these, such that $\kappa$ proxies the scale factor. Here, we effectively vary the scale by 20\% instead of 40\%.}

\begin{figure}[h!]
    \centering
    \includegraphics[scale=0.55]{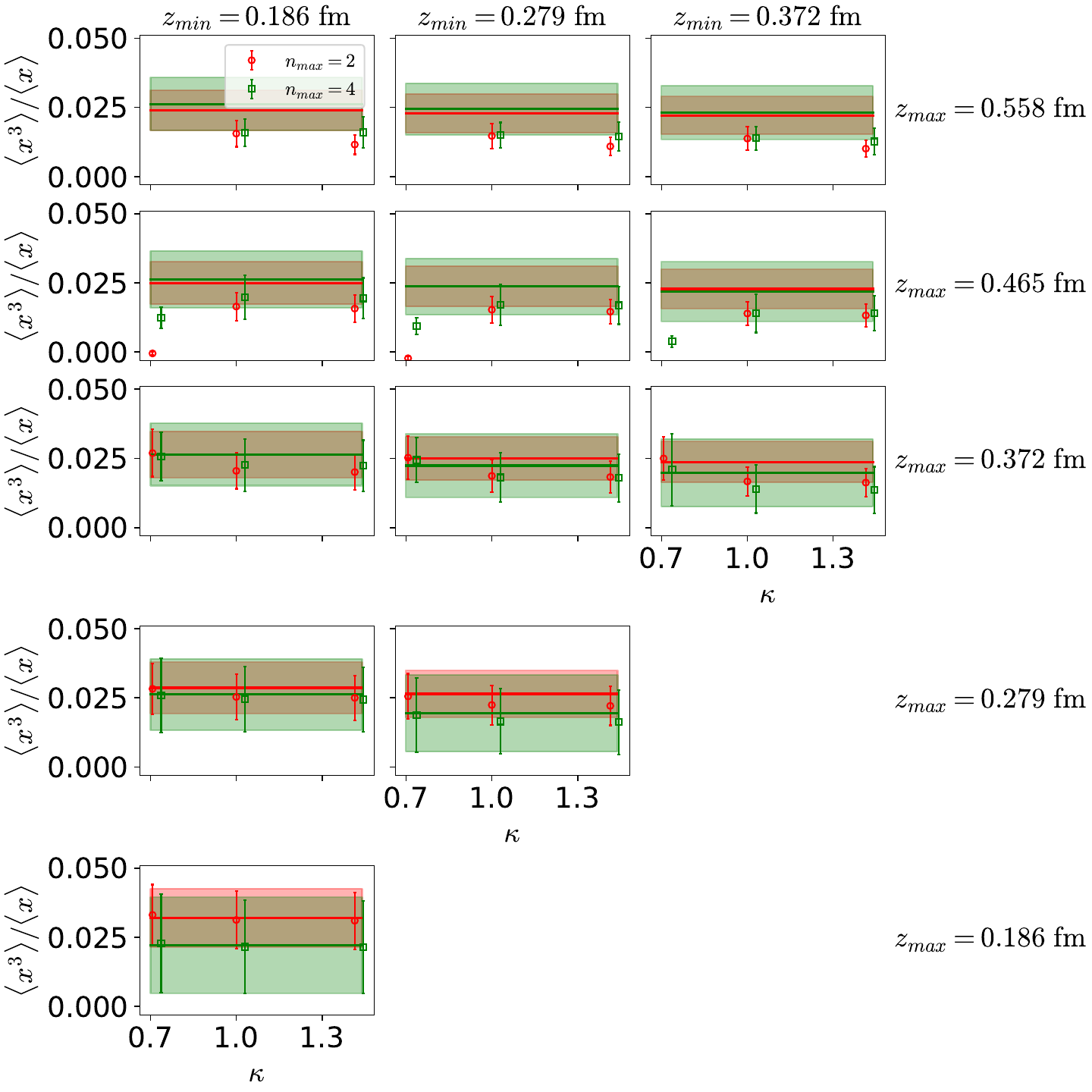}
    %\vspace*{-0.5cm}
    \caption[A comparison of moments extracted for different $n_{max}$ and $(z_{min},z_{max})$ combinations and evolved from initial scale $\mu_0$ to 2 GeV]{A comparison of moments extracted for different systematics and evolved from initial scale $\mu_0$ to 2 GeV, and $\kappa = \frac{\mu_0^2 z^2 e^{2\gamma_E}}{4}$. OPE sum truncation order is indicated for red circles for $n_{max}=2$, and green squares for $n_{max}=4$, and fixed-order results are shown as bands of corresponding colors. Plots are organized in ascending order of $z_{min}/a = 2,\,3,\,4$ from left to right and in descending order of $z_{max}/a =6,\,4,\,2$ from top to bottom.}
    \label{fig:evo_moments}
\end{figure}

In Fig.~\ref{fig:evo_moments}, we present the resulting ratio $\langle x^3\rangle_g/\langle x\rangle_g$ evolved to $\mu = 2$~GeV as a function of $\kappa$, for different fit windows and truncation orders $n_{max}=2,4$;  $n_{max}=6$ is not shown due to the large statistical errors. Additionally, for $\kappa=\frac{1}{\sqrt{2}}$, the scale $\mu_0$ falls below the perturbative regime at large $z$, approaching the Landau pole of $\alpha_s(\mu)$. This leads to either unphysically small values (see, e.g., $z_{min}=2a-4a$ and $z_{max}=5a$) or a non-convergent fit (see, e.g., $z_{min}=2a-4a$ and $z_{max}=6a$). The fixed-order result of Sec.~\ref{sec:fixed_order} is shown as a band for comparison. Similarly to the previous figures, the plots are organized from left to right in ascending order of $z_{min}$ and from top to bottom in descending order of $z_{max}$. Several observations can be made: First, for moderate values of $\kappa$ around 1, the evolved results are statistically compatible with the fixed-order determination. Second, for larger values of $z_{max}$ and small $\kappa$, deviations become visible. In this region, the corresponding $\mu_0$ is small, leading to larger values of $\alpha_s$ and, therefore, reduced perturbative control. This behavior is consistent with expectations from the logarithmic structure of the matching. Third, for smaller $z_{max}$, where the OPE is expected to be more reliable, the $\kappa$ dependence remains mild over the full variation range. We note that in this plot we neglect mixing with the quark-singlet contribution, as the focus is on the effect of RG evolution. 

\begin{figure}[h!]
    \centering
    \includegraphics[scale=0.65]{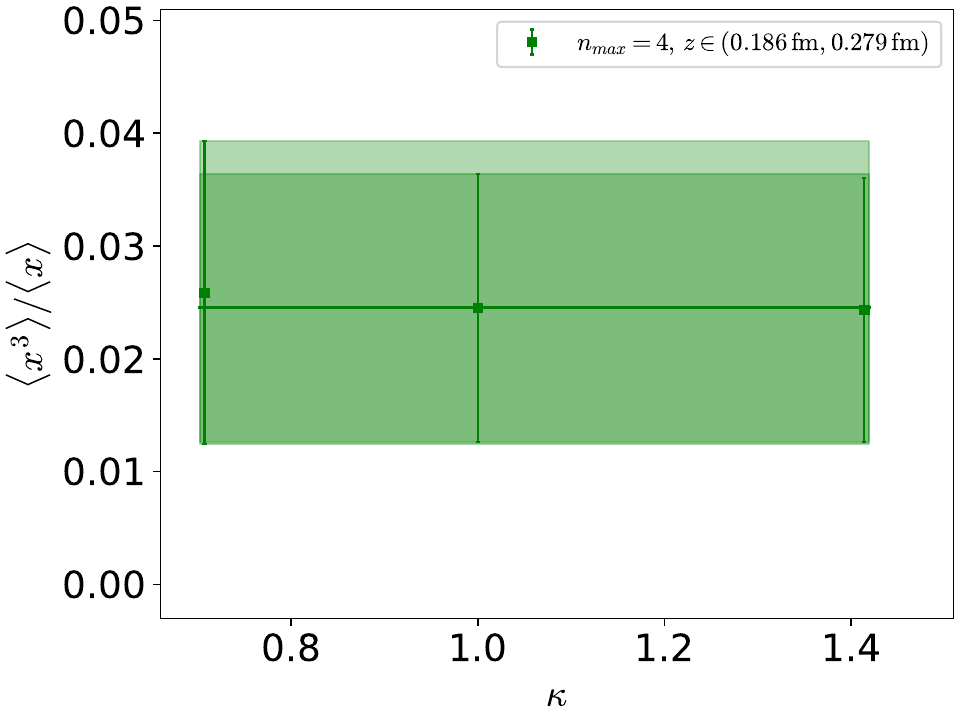}
    %\vspace*{-0.5cm}
    \caption{Comparison of ratios $\langle x^3 \rangle_g/\langle x \rangle_g$ for gluon only at final choice of $z$-values and OPE expansion order incorporating scale evolution. The data are presented at $\kappa=\frac{1}{\sqrt{2}},\, 1,$ and $\sqrt{2}$. We include a band covering the uncertainties associated with extracted value at $\kappa=1$ and a lighter band including additional theoretical error determined from varying $\kappa$.}
    \label{fig:evo_moments_g_only_final}
\end{figure}

For clarity of presentation we select the RG-evolved results for $n_{max}=4$ and $(z_{min},\,z_{max}) = (2a,\,3a)$ as shown in Fig.~\ref{fig:evo_moments_g_only_final} at $\kappa = 1/\sqrt{2},\,1,\,\sqrt{2}$. The band is taken at $\kappa=1$, while the variation under $\kappa \to 1/\sqrt{2},\,\sqrt{2}$ provides an estimate of the perturbative uncertainty associated with matching-scale variation, in addition to the statistical uncertainties.

%\newpage
\subsection{Impact of Mixing with Quark-Singlet Contribution}
\label{sec:mixing}

In the extraction of gluon Mellin moments, the perturbative matching receives contributions from mixing with the quark-singlet operator. As discussed in Sec.~\ref{sec:methods_A}, this contribution enters the short-distance expansion at $\mathcal{O}(\alpha_s)$ and, in principle, should be included in the analysis. However, given the current statistical precision of the lattice QCD data for both the gluon and quark-singlet contributions, both of which require the calculation of disconnected diagrams, there is limited sensitivity to simultaneously constrain both gluon and quark-singlet contributions from purely lattice QCD data. Consequently, the main analysis presented in previous sections neglects the mixing term in Eq.~\eqref{eq:ME_OPE_matching}. Here, we provide a qualitative estimate of its impact by incorporating phenomenological information on $\langle x^3 \rangle_S$ from the unpolarized quark-singlet PDFs of the NNPDF4.0 global analysis~\cite{Ball_2022}. For completeness, we provide the values for the gluon and quark singlet moments that are obtained for this analysis: $\langle x^m \rangle_g = \{0.400(3),\ 0.0112(4),\ 0.0016(2),\ 0.00038(7)\}$, $\langle x^m \rangle_S = \{0.600(3),\ 0.0468(3),\ 0.0099(1),\ 0.0031(1)\}$, for $m=1,3,5,7$, respectively.
Since the mixing between the gluon and quark-singlet contributions enters order-by-order in the OPE formalism, we obtain the ratio $\langle x^3 \rangle_S / \langle x \rangle_g$ at 2 GeV sampled from proton PDF replicas accessed via the LHAPDF database~\cite{Buckley_2015}. Following the procedure of extracting $\langle x^3 \rangle_g$ from lattice QCD data, we impose the same truncation order in the OPE, namely $n_{max}= 2, 4, 6$. 

\begin{figure}[h!]
    \centering
    \includegraphics[scale=0.55]{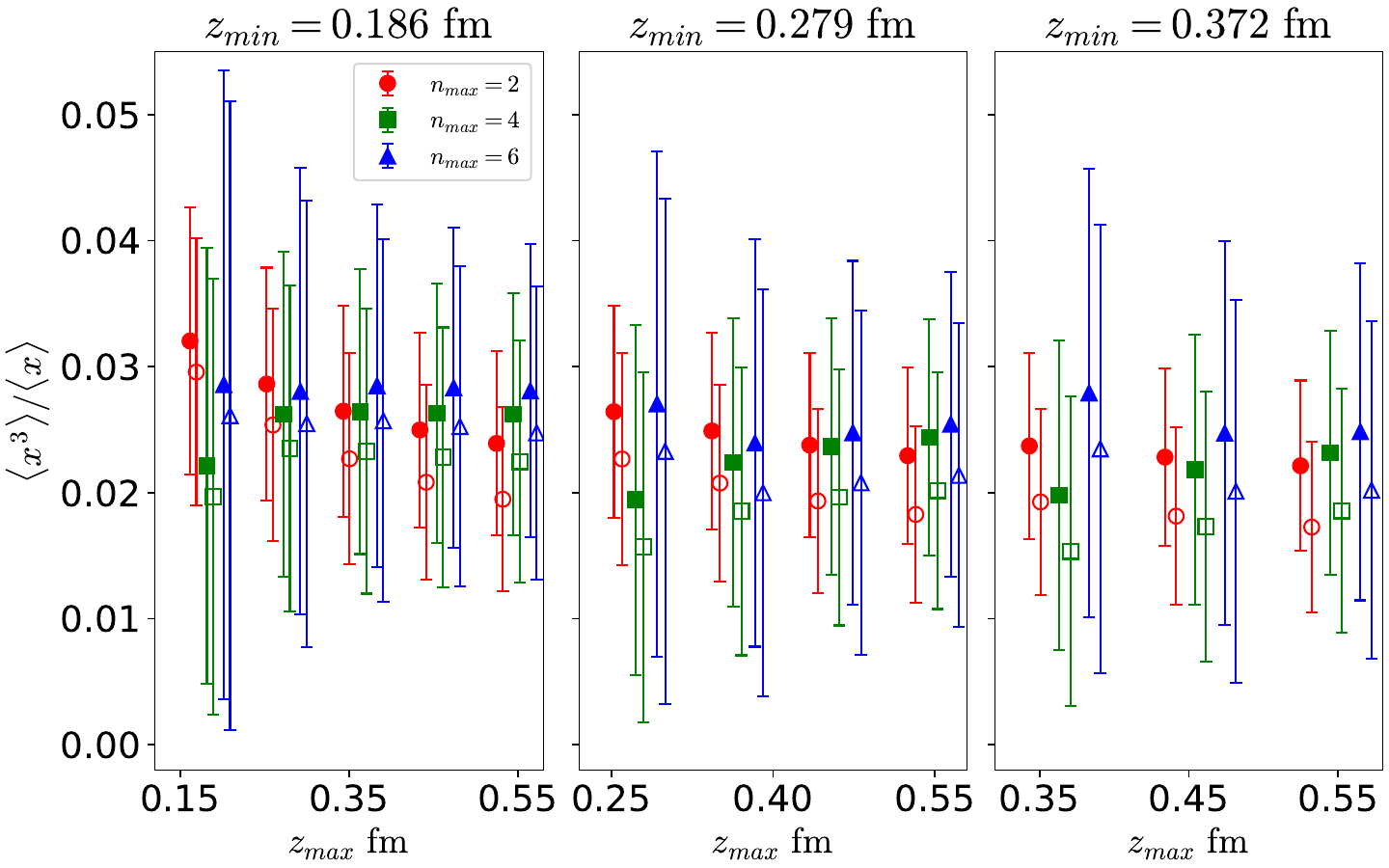}
    %\vspace*{-0.5cm}
    \caption{Fixed-order results for $\langle x^3 \rangle_g/\langle x \rangle_g$ when incorporating the mixing with the quark-singlet contribution. Filled points are the same as Fig.~\ref{fig:no_evo_moments}. Unfilled points utilize quark-singlet data from Ref.~\cite{Ball_2022}. Data points incorporating the mixing are offset for visibility.}
    \label{fig:no_evo_moments_singlet}
\end{figure}

Fig.~\ref{fig:no_evo_moments_singlet} presents the fixed-order extraction of the ratio $\langle x^3\rangle_g/\langle x\rangle_g$ when the quark-singlet contribution is included in the perturbative matching (open symbols), compared with the results obtained without the mixing effects taken for Fig.~\ref{fig:no_evo_moments} (filled symbols). The layout of the plot follows that of Fig.~\ref{fig:no_evo_moments}: the three panels correspond to increasing values of $z_{min}$ from left to right, while the horizontal axis in each panel indicates the choice of $z_{max}$. Different symbols represent different truncation orders, $n_{max}$, of the OPE.
The main effect of including quark-singlet mixing is a small but systematic downward shift in the extracted values of $\langle x^3\rangle_g/\langle x\rangle_g$ across all fit windows and truncation orders. The magnitude of the shift is relatively uniform and does not alter the qualitative behavior observed in the analysis when the mixing is ignored. In particular: \textbf{(i)} the dependence on $z_{min}$ remains mild; \textbf{(ii)} the variation with $z_{max}$ continues to exhibit stability within uncertainties; and \textbf{(iii)} the compatibility among truncation orders $n_{max}=2,4,6$ is preserved, indicating that the convergence pattern of the OPE remains unaffected by the inclusion of the mixing term. The modest size of the observed shift is consistent with expectations from perturbative quark–gluon mixing, where the quark-singlet contribution enters as a sub-leading correction, and is suppressed relative to the dominant gluonic term. Within the current statistical precision, the mixing effect remains smaller than or comparable to the fit uncertainties. These findings therefore suggest that neglecting the quark-singlet contribution does not introduce a significant systematic bias at the present level of accuracy, although a more complete treatment of the coupled gluon and quark-singlet system will become increasingly important in future high-precision calculations.

\subsection{Final Results}

\begin{figure}[h!]
    \centering
    \includegraphics[scale=0.65]{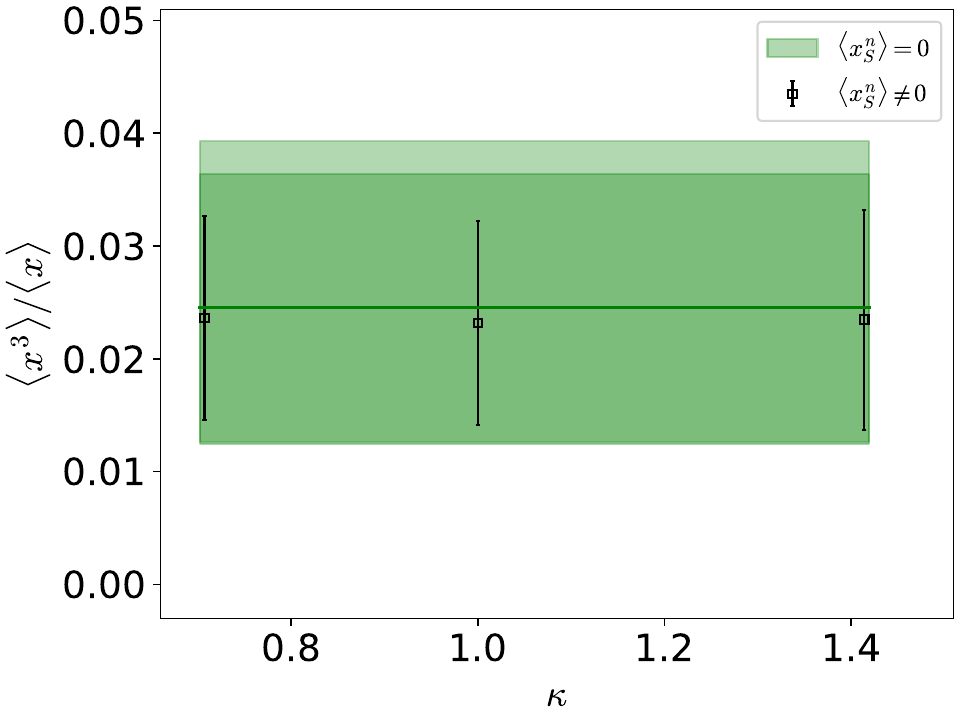}
    %\vspace*{-0.5cm}
    \caption{A comparison of ratios $\langle x^3 \rangle_g/\langle x \rangle_g$ including (data points) and neglecting (bands) the quark-singlet contribution. Scale evolution is incorporated in both cases, and the results are presented at $\kappa=\frac{1}{\sqrt{2}},\, 1,$ and $\sqrt{2}$. }
    \label{fig:evo_moments_singlet_final}
\end{figure}

Based on the conclusions reported in previous sections, we take as our final result the determination shown in Fig.~\ref{fig:no_evo_moments_singlet}, which incorporates scale evolution. In particular, we select the analysis with $(z_{min},z_{max})=(2a,3a)$ as our preferred window, and truncation order $n_{max}=4$, which provides an optimal choice between perturbative control, stability of the extracted moments, and statistical precision. The corresponding evolved results accounting for mixing effects are presented as points in Fig.~\ref{fig:evo_moments_singlet_final} for  $\kappa=1/\sqrt{2},\,1,\,\sqrt{2}$. The shaded band corresponds to the final lattice-only determination obtained from the analysis without quark-singlet mixing of Eq.~\eqref{eq:final_value}. The central value is taken from the $\kappa=1$ analysis, corresponding to the natural choice $\mu_0\sim1/z$, while the variation under $\kappa\rightarrow1/\sqrt{2}$ and $\kappa\rightarrow\sqrt{2}$ is used to estimate the associated perturbative systematic uncertainty. 

Our final result obtained from purely lattice QCD data, that is, neglecting the mixing effect, is 
\begin{equation}
\label{eq:final_value}
\frac{\langle x^3\rangle_g}{\langle x\rangle_g}
\Big|_{\mu=2~{\rm GeV}} =
0.0250 \pm 0.0120^{+0.0029}_{-0.0002}\,,
\end{equation}
where the first uncertainty is statistical and the asymmetric second uncertainty reflects the variation under $\kappa$ as described above. We compare our results with the corresponding determinations from global QCD analyses at a scale of 2 GeV. From the NNPDF4.0 analysis~\cite{Ball_2022} at NNLO, one obtains $\frac{\langle x^3 \rangle_g}{\langle x \rangle_g} = 0.0281(11)$ with $\langle x \rangle=0.400(3)$. Additionally, a recent analysis by the JAM collaboration done at NLO gives $\frac{\langle x^3 \rangle_g}{\langle x \rangle_g} = 0.0317(17)$ with $\langle x \rangle=0.433(5)$~\cite{Cocuzza:2025qvf, cocuzza2026stabilitypartondistributionshigh}. Both phenomenological determinations are consistent with our lattice QCD result within uncertainties. At the present level of precision, no significant tension is observed between the lattice QCD extraction and global analyses. The results are summarized in Table~\ref{tab:comparison_gluon_moments}.
%If comparing to NNPDF4.0 <x^5>/<x> use 0.0040(4) and JAM 0.0018(5)
For completeness, the table also includes the corresponding extraction obtained when incorporating the mixing with the quark-singlet contribution discussed in Sec.~\ref{sec:mixing}, enabling a direct comparison between the two analyses. It is interesting to note that the result incorporating mixing leads to reduced uncertainties. This reduction is driven in part by the comparatively precise phenomenological constraints on the quark-singlet moments. However, the stability of the fit and the moderate shift in the central values indicate that the phenomenological singlet moments are broadly consistent with those favored by the lattice matrix elements. Were there significant tension between the two inputs, one would generally expect larger residual uncertainties or noticeable fit instabilities.

\begin{table}[h!]
\centering
\renewcommand{\arraystretch}{2}
\begin{tabular}{|l|c|c|}
\hline
\textbf{Analysis} 
& $\displaystyle {\langle x^3\rangle_g}/{\langle x\rangle_g}$ 
& $\displaystyle \langle x\rangle_g$ \\
\hline
Lattice QCD, no mixing incorporated (this work) 
& $\qquad\,$ $0.0250(120)^{+0.0029}_{-0.0002}$ $\quad$
& --- \\
Lattice QCD, mixing incorporated (this work) 
& $\qquad\,$ $0.0232(90)^{+0.0010}_{-0.0004}$ $\quad$
& --- \\
%$0.0266(98)^{+0.0010}_{-0.0003}$
NNPDF4.0~\cite{Ball_2022} 
&$\quad$ $0.0281(11)$ $\quad\quad\,\,\,$
& $0.400(3)$  \\

JAM (NLO)~\cite{Cocuzza:2025qvf,cocuzza2026stabilitypartondistributionshigh} 
&$\quad$ $0.0317(17)$ $\quad\quad\,\,\,$
& $0.400(6)$ \\
\hline
\end{tabular}
\caption{Comparison of the gluon moment ratio 
$\langle x^3\rangle_g/\langle x\rangle_g$ 
at $\mu=2$~GeV from lattice QCD and global analyses. The lattice QCD data use $(z_{min},z_{max})=(2a,3a)$, and $n_{max}=4$. The asymmetric second uncertainty reflects the variation under $\kappa$ in the RG evolution.}
\label{tab:comparison_gluon_moments}
\end{table}

\section{Summary and Outlook}
\label{sec:conclusions}

In this work, we presented a lattice QCD determination of higher Mellin moments of the unpolarized gluon PDF in the proton using matrix elements of nonlocal gluon operators within the operator product expansion. By analyzing the short-distance behavior of the reduced Ioffe-time distribution, we extracted the ratio of $\langle x^3\rangle_g / \langle x \rangle_g$; higher Mellin moments are too noisy and magnitude-suppressed to be able to constrain them. Our analysis included a study of systematic effects arising from truncation of the OPE, the range of spatial separations included in the fit, and the treatment of quark-singlet mixing contributions. We further investigated the stability of the extracted moments under DGLAP evolution and assessed the uncertainties associated with varying the matching scale.

We found that the extracted values of $\langle x^3\rangle_g/\langle x \rangle_g$ from the various analyses remain stable within uncertainties under variations of the fit window, truncation order, and perturbative matching scale. The RG-evolved results are consistent with the fixed-order analysis at perturbative scales within the expected region of validity of the short-distance expansion. We also investigated the impact of quark-singlet mixing using phenomenological input from global analyses and found that its contribution is relatively small compared to the present statistical uncertainties. Our final result for $\langle x^3\rangle_g/\langle x \rangle_g$ at $\mu=2$ GeV is consistent within uncertainties with recent determinations from global QCD analyses, as reported in Table~\ref{tab:comparison_gluon_moments}.

This work demonstrates that Mellin moments of the gluon PDF can be accessed directly from nonlocal lattice QCD matrix elements without requiring a full reconstruction of the $x$ dependence. In particular, determining $\langle x^3\rangle_g$ is currently extremely challenging, if not impossible, using traditional approaches based on local operators, due to the rapid growth of operator complexity, mixing patterns, and statistical noise. The present study shows that combining nonlocal operators with the short-distance OPE offers an alternative strategy for obtaining meaningful estimates of higher moments directly from lattice QCD data, even though current precision remains limited. At the same time, the analysis highlights the challenges associated with limited Ioffe-time coverage, perturbative truncation effects, and the increasing statistical noise affecting higher moments. While the extraction of $\langle x^5\rangle_g/\langle x\rangle_g$ and higher moments remains unconstrained, the methodology used here provides a framework for future high-precision investigations of gluon structure from lattice QCD.

Several extensions of this work are natural to pursue. A next step for the same ensemble would be to increase statistics and larger hadron boosts, which will further extend the accessible Ioffe-time range and improve sensitivity to higher moments. In addition, calculations at multiple lattice spacings would enable a continuum extrapolation and improved quantification of discretization effects. Finally, the direct lattice QCD determination of the quark-singlet contribution eliminates the need for phenomenological input. More broadly, the methods presented here can be generalized to polarized gluon distributions and gluonic structure in mesons, providing additional opportunities for lattice QCD to contribute to the precision hadron-structure program relevant for the future Electron-Ion Collider and other experimental facilities.

\begin{acknowledgements}

J.D received support from Argonne National Laboratory under the contract ``Pion and Kaon Form Factors using Lattice QCD''.
K.~C.\ is supported by the National Science Centre (Poland) grant OPUS No.\ 2021/43/B/ST2/00497. 
M.~C. acknowledges financial support by the U.S. Department of Energy, Office of Nuclear Physics,  under Grant No.\ DE-SC0025218. The work of YZ is supported by the U.S. Department of Energy, Office of Science, Office of Nuclear Physics through Contract No.~DE-AC02-06CH11357, and the Early Career Award through Contract No.~DE-SCL0000017. 
The authors acknowledge partial support from the U.S. Department of Energy, Office of Science, Office of Nuclear Physics, under the umbrella of the Quark-Gluon Tomography (QGT) Topical Collaboration, with Award DE-SC0023646.

This research includes calculations carried out on HPC resources supported in part
by the National Science Foundation through major research instrumentation grant number 1625061 and by the US Army Research Laboratory under contract number W911NF-16-2-0189. This research used resources of the Oak Ridge Leadership Computing Facility, which is a DOE Office of Science User Facility supported under Contract DEAC05-00OR22725.
This research used resources of the National Energy Research
Scientific Computing Center, a DOE Office of Science User Facility
using NERSC award ALCC-ERCAP0030652.
The gauge configurations have been generated by the Extended Twisted Mass Collaboration on the KNL (A2) Partition of Marconi at CINECA, through the Prace project Pra13\_3304 ``SIMPHYS".
Inversions were performed using the DD-$\alpha$AMG solver~\cite{Frommer:2013fsa} with twisted mass support~\cite{Alexandrou:2016izb}. 
\end{acknowledgements}

\bibliographystyle{h-physrev}
\bibliography{references.bib}

\end{document}